\documentclass{article}  
\usepackage{jamaica04}
\usepackage{graphicx}
\frompage{000} \topage{000}                                              
\def\rightmark{Azimuthal anisotropy in high-energy heavy-ion}
\def\leftmark{ShinIchi Esumi}
 
\title{Azimuthal anisotropy in high-energy heavy-ion \\
collisions at RHIC energies }
\authors{
{ShinIchi Esumi}\\[2.812mm]
{\normalsize
Institute of Physics, University of Tsukuba, \\
Tsukuba, Ibaraki 305, Japan\\[0.2ex] 
}}
 
\abstract{Directed and elliptic event anisotropy parameters measured
in the experiments at relativistic heavy-ion collider are presented. 
The possible origin of the measured elliptic anisotropy parameter 
$v_2$ and its sensitivity to the early phase of the high-energy 
heavy-ion collisions are discussed.}

 
\begin{document}
 
\maketitle
\setcounter{page}{1}

\section{charged particle $v_1$, $v_2$}\label{sec1}

\begin{figure}[ht]
\includegraphics[width=120mm]{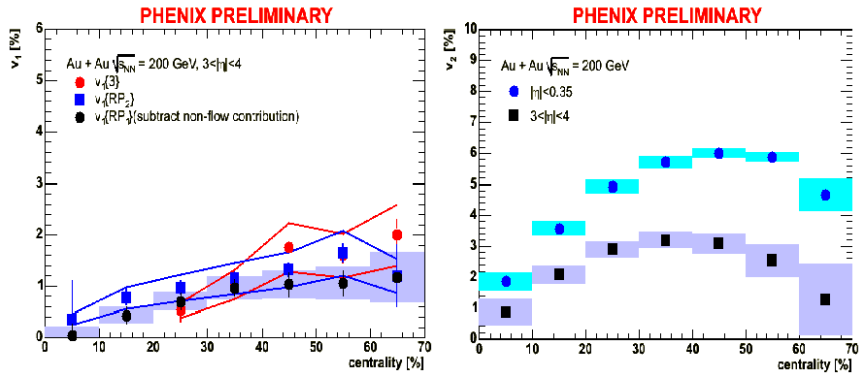}
\caption[]{The charged particle $v_1$ (3 $<$ $|\eta|$ $<$ 4) 
as a function centrality is shown in the left figure. The 
charged particle $v_2$ ($|\eta|$ $<$ 0.35 and 3 $<$ $|\eta|$ 
$<$ 4) are shown in the right figure.}
\label{fig1}
\end{figure}

The charged particle $v_1$ and $v_2$ in the forward rapidity 
(3 $<$ $|\eta|$ $<$ 4) are presented as a function of centrality 
in Fig.\ref{fig1}. 
The 3 different extraction methods are tested to extract $v_1$.
3 particles selected from 3 rapidity intervals (two particles from
the forward and backward rapidities and one particle from the mid-
rapidity with the 2nd harmonics are used), the magnitude of $v_1$
in the forward rapidity is calculated with a knowledge of $v_2$ in 
the mid-rapidity.\cite{bog1} 
The 2nd method replaces the 3rd particle with
the reaction plane (2nd moment) from the mid-rapidity. The 3rd 
method uses the reaction plane from the opposite rapidity, where 
the non-flow effect from the momentum conservation is expected, 
the expected non-flow contribution is subtracted here. The first
two methods is expected to be less affected by such a effect. 
The right panel in Fig.\ref{fig1} shows the $v_2$ measured in 
the forward rapidity compared with the mid-rapidity, both are 
the $p_T$ integrated $v_2$, it is interesting to note that there
seems to have slight different centrality dependence for two 
rapidity slices. 
 
\section{Identified particle $v_2$}\label{sec2}

\begin{figure}[ht]
\includegraphics[width=120mm]{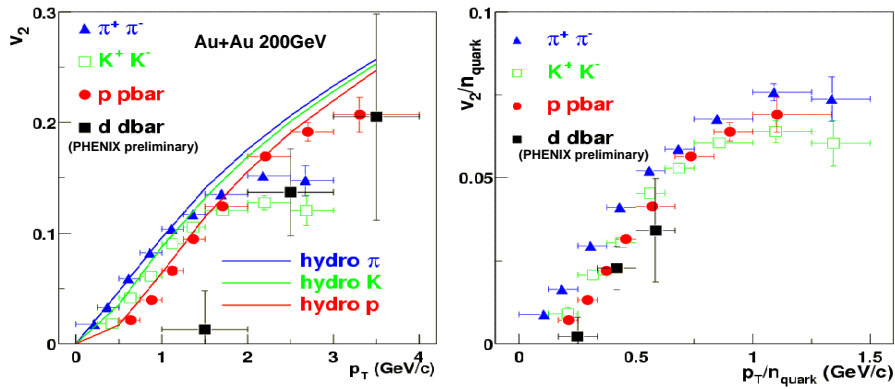}
\caption[]{The identified particle $v_2$ ($|\eta|$ $<$ 0.35) 
for $\pi^{+-}$, $K^{+-}$, $p+pbar$ and $d+dbar$ are shown 
in the left figure. The right figure shows the same for the 
number of quark scaled in both $v_2$ and $p_T$.}
\label{fig2}
\end{figure}
 
The identified particle $v_2$ in the mid-rapidity is shown 
in the Fig.\ref{fig2}\cite{adl1} and compared to a hydro
model.\cite{huo1}
The reaction plane is defined in the 
forward and backward rapidities and the particle identification
is given by the time of flight detector. The right panel 
shows the number of quark scaled $v_2$, the agreement between 
different particle species might tells us the $v_2$ is already 
generated during the partonic phase (before the hadrons are 
formed), but there seems to be some remaining mass ordering 
which might come from the later stage (hadronic flow), although
the difference between pion $v_2$ and the other hadrons is well
described by the feed-down effects.\cite{gre2,don1}

\begin{figure}[ht]
\includegraphics[width=120mm]{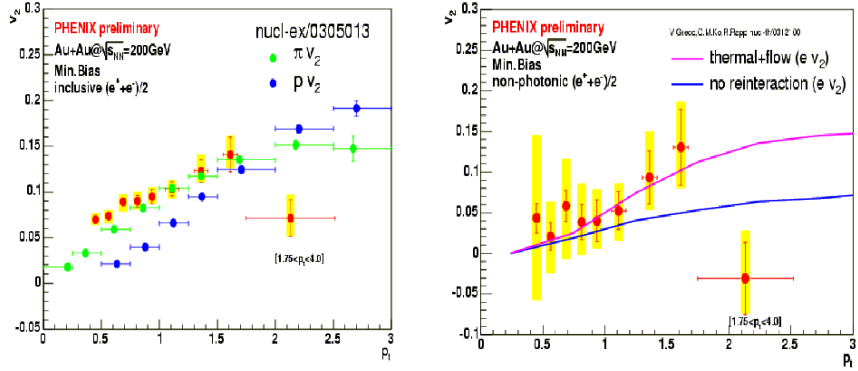}
\caption[]{The inclusive electron/positron $v_2$ is shown 
in the left figure. The right figure shows the $v_2$ of 
subtracted charmed electron contribution.}
\label{fig3}
\end{figure}

The electron and positron $v_2$ are measured as shown in 
the left panel of the Fig.\ref{fig3} in order to 
extract the charm quark $v_2$, since the single electron 
momentum distribution is measured to be dominated by the 
semi-leptonic decay contribution of the open charm at high
$p_T$.\cite{sea1} 
The electron and positron are identified by the 
ring imaging Cherenkov detector and the electro magnetic 
calorimeter. Using the relative yield of open charm in the 
inclusive electron momentum spectrum and the measured 
$v_2$ of the dominant $\pi^0$ contribution, the $v_2$ of 
charmed electron are extracted as shown in right panel in
the Fig.\ref{fig3} and compared with a model.\cite{gre1}
The present statistics does not allow 
us to conclude on the charmed quark $v_2$.

\begin{figure}[ht]
\includegraphics[width=120mm]{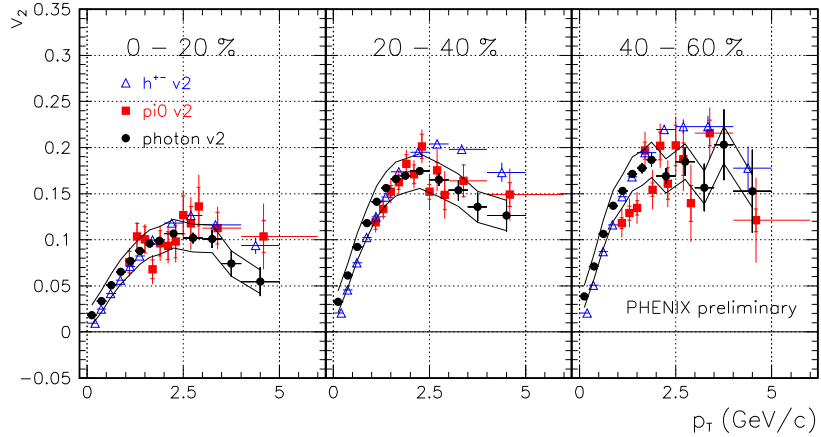}
\caption[]{The inclusive photon $v_2$ compared to the $\pi^0$
and $h^{+-}$ $v_2$ as a function of $p_T$ for 3 centrality bins.}
\label{fig4}
\end{figure}
 
The inclusive photon $v_2$ are measured and compared with 
$\pi^0$ $v_2$ as shown in the Fig.\ref{fig4}.\cite{mas1}
The photon and $\pi^0$ are measured and 
reconstructed with the electro magnetic calorimeter. 
Using the direct photon enhancement with respect to the 
$\pi^0$ decay contribution, which seems to be consistent 
with the binary collision scaled production of the direct 
photon,\cite{jus1} 
the direct photon $v_2$ would be extracted in the 
future run with much higher statistics. 

\section{Origin of $v_2$ and non-flow}\label{sec3}
 
\begin{figure}[ht]
\includegraphics[width=120mm]{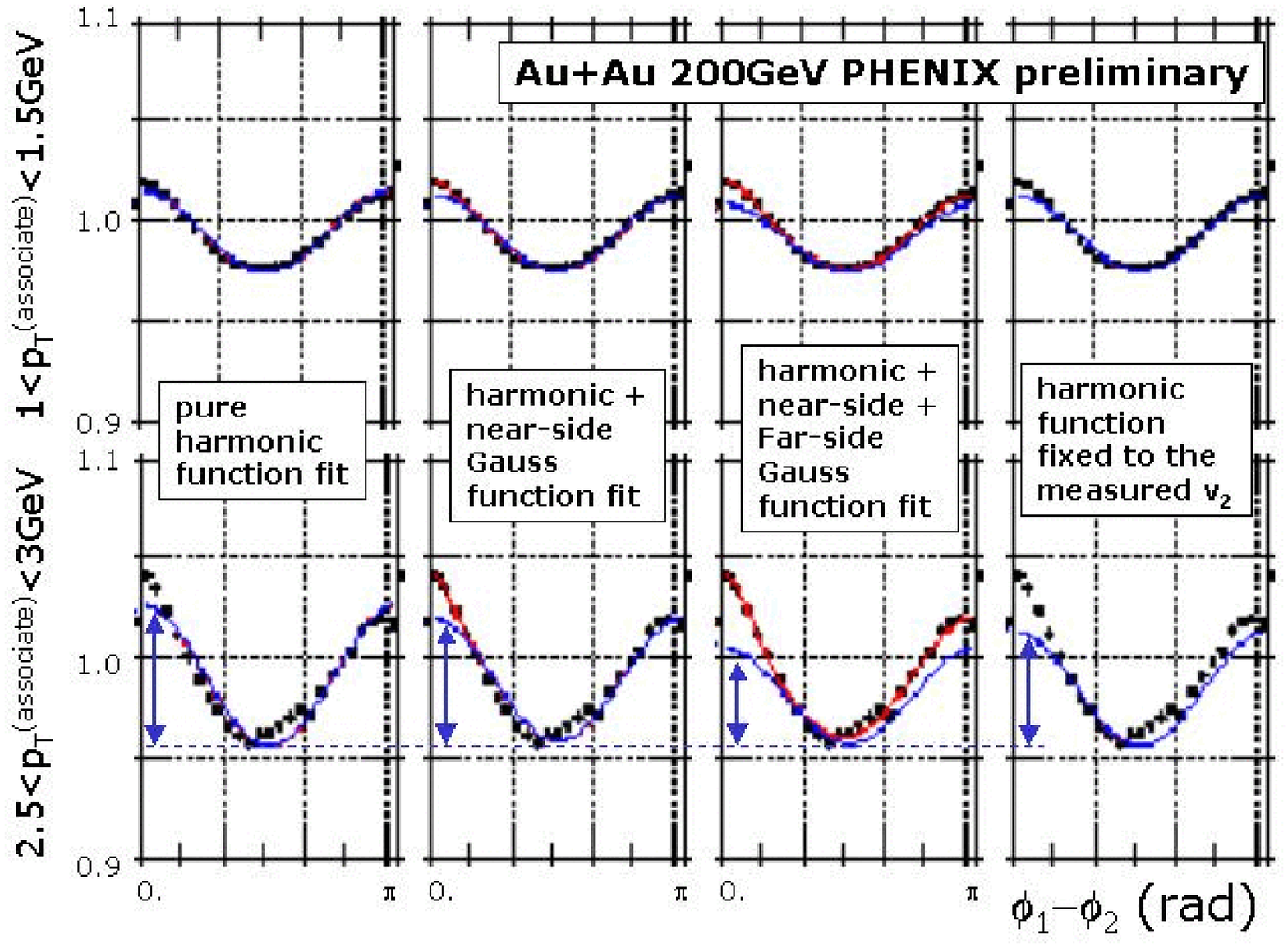}
\caption[]{The two particle correlation for two $p_T$ bins
(top and bottom) and different fitting functions (left to
right) to extract $v_2$ are shown.} 
\label{fig5}
\end{figure}
 
The assorted correlation functions are shown with one 
particle in the full $p_T$ reference and another in the 
$p_T$ window described in the Fig.\ref{fig5}. From the
left to right, 
(1) pure harmonic function, 
(2) pure harmonic + near side Gaussian function, 
(3) pure harmonic + near/away side Gaussian function, 
(4) pure harmonic where the magnitude of $v_2$ fixed 
by the measurement, only the fitting functions are 
different from the left to right, 
while the data are the same (top for the 
lower $p_T$, and bottom for the higher $p_T$ windows). 
It is clearly demonstrated that the extracted magnitude 
of $v_2$ from the pair azimuthal correlation depends 
on what type of function is used to fit the measured 
correlation in order to remove the non-flow contribution. 
The comparison shows that the measured 
correlation suggests that there is a need to have a 
rather wide away side Gaussian-like (although it does 
not have to be a Gaussian shape) contribution. 

\begin{figure}[ht]
\includegraphics[width=120mm]{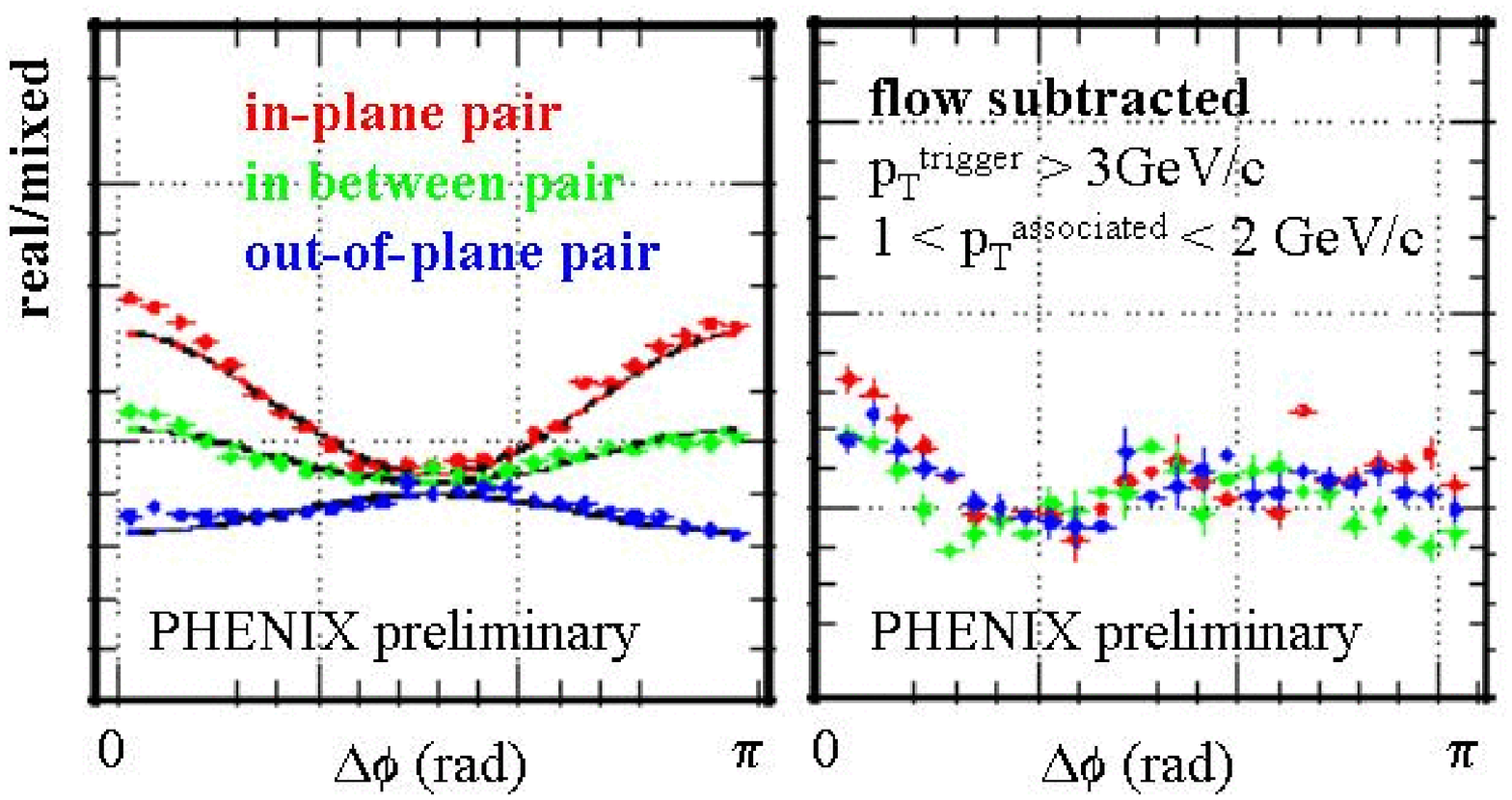}
\caption[]{The two particle correlation with respect to 
the different reaction plane orientation. The fitted 
function is given by the measured $v_2$ with respect to 
the reaction plane and reaction plane resolution.}
\label{fig6}
\end{figure}
 
The Fig.\ref{fig6} shows two particle azimuthal correlation
with trigger particle in- or out-of- reaction plane. 
The fitted function has an fixed shape which is given by
an independent measurement of $v_2$ with respect to the 
reaction plane defined in the forward rapidity and the 
reaction plane resolution.\cite{jan1} 
The correlation functions are
mostly described except the clear near side jet contribution 
seen at around 0 degree. However looking at more precisely
the difference between in-plane and out-of-plane as shown 
in the right panel of the Fig.\ref{fig6}, where the expected 
flow contribution is subtracted, the enhancement of in-plane 
correlation over the expected flow contribution seems to 
be larger compared to one from the out-of-plane correlation 
for the both near side (0 degree) and away side (180 degrees). 
This might be an indication of the coalescence of the 
jet-fragmented quarks, which depends on the orientation 
with respect to the reaction plane and this could be one 
of the source of $v_2$ on top of the initial pressure gradient.

\section{Conclusions}\label{sec4}

The charged particle $v_1$ and $v_2$ as a function of centrality
at forward- and mid- rapidities. The measured identified hadron 
$v_2$ shows rather good agreement with quark number scaling by 
taking into account the feed-down effect on pions, which tells
us the $v_2$ might have been formed during the partonic phase,
however there could be another indication of the flow in later 
stage after hadronization in the remaining mass ordering of $v_2$.
The inclusive electron and photon $v_2$ measurements in the future 
run with large statistics will provide the charm quark and direct
photon $v_2$. Combined analysis with two particle correlation
and reaction plane orientation have opened a new dimension of the
event anisotropy and jet tomography analysis and have an 
indication of another origin of the elliptic flow.

\vfill\eject
\end{document}